\documentclass[letterpaper]{sfchem}
\usepackage{graphicx}

\def\k13{\mbox{$\kappa_{\mbox{\tiny 1.3mm}}$}}

\def\rout{\mbox{$R_{\mbox{\tiny out}}$}}
\def\kms{\rm{km\, s^{-1}}}

\def\cmg{\rm{cm^{2} \, g^{-1}}}
\def\as{\mbox{$a_{\mbox{\tiny s}}$}}
\def\nhd{\mbox{$n_{\mbox{\tiny H}_{2}}$}}
\def\>{$>$}
\def\<{$<$}

\def\ltsima{$\; \buildrel < \over \sim \;$}
\def\simlt{\lower.5ex\hbox{\ltsima}}
\def\gtsima{$\; \buildrel > \over \sim \;$}
\def\simgt{\lower.5ex\hbox{\gtsima}}
\def\arcsec{\hbox{$^{\prime\prime}$}}

\def\h2{\mbox{$_{\mbox{\tiny H2}}$}}

\begin{document}

\title{Submillimeter Studies of Prestellar Cores and Protostars: 
~~~~~~~~~~Probing the Initial Conditions for Protostellar Collapse}

\author{Philippe\ Andr\'e\inst{1} \and Jeroen Bouwman\inst{1} \and 
Arnaud Belloche\inst{1,2} \and Patrick Hennebelle\inst{2,3} } 
  \institute{CEA Saclay, Service d'Astrophysique, 
  Orme des Merisiers, B\^at.~709, F-91191 Gif-sur-Yvette Cedex,  
  France \and Laboratoire de Radioastronomie, ENS, 24 rue Lhomond, 
F-75231 Paris Cedex 05, France \and Department of Physics \& Astronomy, 
Cardiff University, Cardiff CF24 3YB, Wales, UK} 
%Short author list here: Surnames only please (no initials)
\authorrunning{Andr\'e, Bouwman, Belloche, and Hennebelle}
%Short title here:
\titlerunning{Submillimeter Studies of Prestellar Cores and Protostars}

\maketitle 

\begin{abstract}
Improving our understanding of the initial conditions and earliest stages 
of protostellar collapse is crucial to gain insight into the origin of 
stellar masses, multiple systems, and protoplanetary disks.
Observationally, there are two complementary approaches to this problem:
(1) studying the structure and kinematics of prestellar cores observed 
prior to protostar formation, and (2) studying the structure of young 
(e.g. Class 0) accreting protostars observed soon after point mass formation.
We discuss recent advances made in this area thanks to 
(sub)millimeter mapping observations with large single-dish telescopes and 
interferometers. 
%Two results have emer-ged. First, 
In particular, we argue that the beginning of 
protostellar collapse is much more violent in cluster-forming clouds than
in regions of distributed star formation. 
%Second, protostars in clusters seem to originate from 
%finite, detached reservoirs of mass, suggesting the IMF is largely 
%determined by pre-collapse cloud fragmentation.
Major breakthroughs are expected in this field from future large 
submillimeter instruments such as Herschel and ALMA. 

\keywords{ISM: structure -- ISM: gravitational collapse -- Stars: protostellar
accretion rates}

\end{abstract}

\section{Introduction}
  
Although the formation of low-mass stars is now reasonably well  
understood in outline (see, e.g., Mannings, Boss, \& Russell 2000 for recent
reviews), several important 
aspects remain poorly known, such as the initial stages of the 
collapse process, the mechanism(s) selecting stellar masses, 
or the formation of multiple systems.  
Some progress on the earliest stages of star formation was achieved 
over the last decade thanks to the use of sensitive receivers  
on large (sub)millimeter radiotelescopes such as JCMT, CSO, and the IRAM 30m. 
Young protostars were identified at the beginning of the main accretion phase 
(Class 0 objects -- Andr\'e, Ward-Thompson, \& Barsony 1993), and 
the starless dense cores extensively studied in NH$_3$ by Myers and 
collaborators (e.g. Benson \& Myers 1989) were found to be characterized by 
flat inner density gradients (Ward-Thompson et al. 1994). 
Direct evidence for infall motions was observed toward a large number of 
Class~0 protostars and prestellar cores (e.g. Gregersen et al. 1997 -- 
see \S ~3 below and Evans, this volume).
Class~0 objects were also found to drive more powerful outflows 
than more evolved (Class~I)
protostars, suggesting a marked decrease of the mass accretion/ejection rates 
in the course of protostellar evolution (Bontemps et al. 1996).  
As advocated by Henriksen, Andr\'e, \& Bontemps (1997), there may be a 
causal relationship between these results, in the sense that the 
accretion/ejection decline during the protostellar phase may be a direct
consequence of the form of the density profile at the prestellar stage.\\
Further studying the detailed properties of prestellar cores and young 
protostars is of prime importance to distinguish between collapse models 
and shed light on the origin of stellar masses. 
Indeed, the effective reservoirs of mass available 
for the formation of individual stars may be largely determined at the prestellar stage, and it is during the main protostellar accretion phase 
that stars accrete some fraction of these reservoirs and build up the masses 
they will have on the zero-age main sequence.

\begin{figure}[ht]
%\resizebox{\hsize}{!}{\includegraphics{/home/storage/pandre/tmp3/fig/l1544_iram%04191.ps}}
\resizebox{\hsize}{!}{\includegraphics{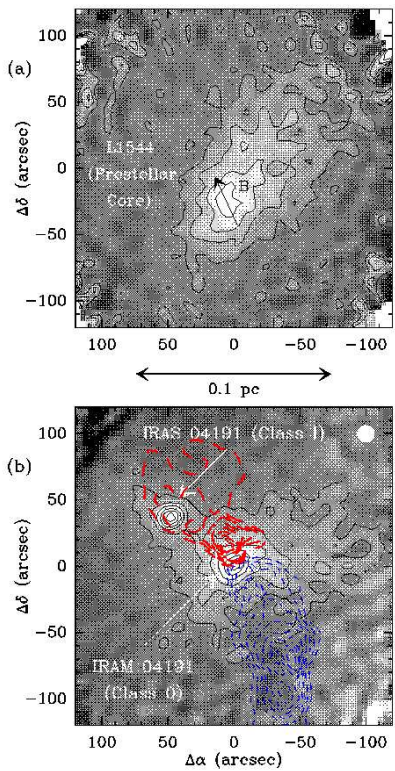}}
\caption{Dust continuum maps of L1544 (a) and IRAM~04191 (b) at 1.3~mm 
taken with the IRAM 30~m telescope and the MPIfR bolometer array 
(from Ward-Thompson et al. 1999 and 
Andr\'e et al. 1999, respectively). Effective resolution: 
13\arcsec ; base contour and contour step: 20~mJy/beam. The direction of 
the magnetic field measured in L1544 with the SCUBA polarimeter on JCMT 
(Ward-Thompson et al. 2000) and the collimated CO(2-1) bipolar flow 
emanating from IRAM~04191 are shown. 
\label{fig-intro}}
\end{figure}
 
After a brief introduction of the theoretical background (\S ~1.1),  
we review recent observational advances concerning the density and 
velocity structure of cloud cores in \S ~2 and \S ~3, respectively. 
We conclude in \S ~4 with a comparison between observations and 
theoretical models.
%\vskip 0.3cm

\subsection{Collapse Initial Conditions: Theory}

The inside-out collapse model of Shu (1977), starting from
a singular isothermal sphere (SIS) or toroid (cf. Li \& Shu 1996, 1997),
is well known and underlies the `standard' picture of isolated, low-mass 
star formation (e.g. Shu, Adams, \& Lizano 1987).\\
Other collapse models exist, however, which adopt different initial 
conditions. In particular, 
Whitworth \& Summers (1985) have shown that there is a two-parameter 
continuum of similarity solutions to the problem of isothermal
spherical collapse. One of the parameters measures how close to hydrostatic 
equilibrium the system is initially, while the other parameter reflects 
how important external compression is in initiating the collapse.
In this continuum, the solutions
proposed by Shu (1977) and Larson (1969)-Penston (1969) represent two
extreme limits. All of the similarity solutions share a universal
evolutionary pattern.  At early times ($t < 0$), a compression wave
(initiated by, e.g., an external disturbance) propagates inward 
%at the sound speed, 
leaving behind it a $\rho(r) \propto r^{-2}$ density profile.  
At $t = 0$, the compression wave reaches the center and a
point mass forms which subsequently grows by accretion.  At later
times ($t > 0$), this wave is reflected into a rarefaction or
expansion wave, propagating outward through the infalling gas 
(at the isothermal sound speed $a_s$), and leaving behind it a 
free-fall $\rho(r) \propto r^{-1.5}$ density distribution. 
Several well-known features of the Shu model
(such as the expansion wave) are thus in fact common to all solutions. 
The various solutions
can be distinguished by the {\it absolute} values of the density and
velocity at $t \sim 0$. In particular, the Shu (1977) solution has
$\rho(r) = (\as^2/2\pi\,G)\ r^{-2}$ and is static ($v = 0$) at $t =
0$, while the Larson-Penston (1969) solution is $\sim 4.4$ times
denser and far from equilibrium ($v \approx -3.3\ \as$).  
During the accretion phase ($t > 0$), the infall envelope is a 
factor $\sim 7$ denser in the Larson-Penston solution. 
%than in the Shu model.
Accordingly, the mass infall rate is also much larger in the Larson-Penston 
case ($\sim 47\, \as^3/G$) than in the Shu case ($\sim \as^3/G$).

In practice, however, protostellar collapse is unlikely to be strictly
self-similar, and the above similarity solutions can 
only be taken as plausible asymptotes. More realistic initial conditions 
than the SIS are provided by the so-called `Bonnor-Ebert' spheres 
(e.g. Bonnor~1956), which represent the equilibrium states for self-gravitating 
isothermal spheres and have a flat density profile in their central $\sim $
Jeans length region. Such spheres are stable for a center-to-edge density 
contrast $< 14.3$ and unstable for a density contrast $> 14.3$ 
(e.g. Bonnor~1956).
Numerical hydrodynamic simulations of cloud collapse starting from such 
initial conditions (e.g. Foster \& Chevalier 1993, Hennebelle et al. 2002) 
find that the Larson-Penston similarity solution is generally a good
approximation near point-mass formation ($t= 0$) {\it at small radii}, 
but that the Shu solution is more adequate at intermediate $t \geq 0$ times,
before the expansion wave reaches the edge of the initial, pre-collapse dense
core. In general, the mass accretion rate is thus expected to be 
time-dependent.

Observationally, it is by comparing the (density and velocity)
structure of prestellar cores such as L1544 (see Fig.~\ref{fig-intro}a)
with the structure of the envelopes surrounding Class~0 protostars 
such as IRAM~04191 (cf. Fig.~\ref{fig-intro}b)   
that one may hope to constrain the initial conditions for collapse and 
to discriminate between the various existing models. 
%In the following, we discuss these two aspects in turn.

\section{Density Structure}
\label{sec-dens}

\subsection{Prestellar cores}

Two main approaches have been used to trace the 
density structure of cloud cores: (1)~mapping the optically thin 
(sub)millimeter continuum {\it emission} from the cold dust contained 
in the cores, and (2) mapping the same cold core dust in {\it absorption} 
against the background infrared emission (originating from warm cloud dust 
or remote stars). 
Mapping the molecular gas component is generally less effective 
as most molecules tend to freeze out onto dust grains in the dense, 
cold inner parts of cloud cores (e.g. Kramer et al. 1999, 
Walmsley et al. 2001, Bacmann et al. 2002, and Bergin 2002).

%The first approach has been used by numerous authors in the last decade
%to discuss the density structure of protostellar envelopes (see \S ~2.3 
%below). 
Ward-Thompson et al. (1994, 1999) and Andr\'e et al. (1996)  
employed the first approach to probe the structure of prestellar cores
(see also Shirley et al. 2000).
Under the simplifying assumption of spatially uniform dust (temperature and
emissivity) properties, they concluded that the radial density profiles of
isolated prestellar cores were {\it not} consistent with the single 
$\rho(r) \propto r^{-2}$ power law of the SIS but were flatter than 
$\rho(r) \propto r^{-1}$ in their inner regions (for $r \leq R_{flat}$), 
and approached $\rho(r) \propto r^{-2}$ only beyond a typical radius
$R_{flat} \sim $~2500--5000~AU. The details of this conclusion have been 
challenged by \cite{Evans01} and \cite{Zuc01} who made the important point 
that starless cores are not strictly isothermal. Accounting for a realistic 
temperature distribution inside the cores (see \S ~2.3 below), these authors
found smaller values for $R_{flat}$ and even claimed that in some cases 
the observed profiles could not be distinguished from a SIS profile. We will
return to this point below (e.g. \S ~2.3).

\begin{figure}[ht!]
%\resizebox{\hsize}{!}{\includegraphics{waterloo_abs.ps}}
\resizebox{\hsize}{!}{\includegraphics{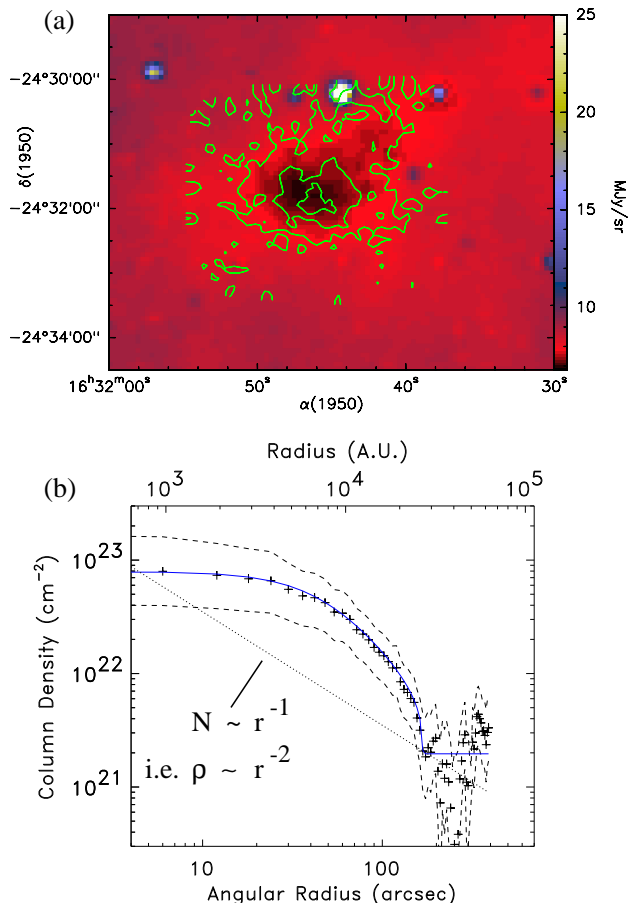}}
\caption{(a) ISOCAM 6.75~$\mu$m absorption image of the prestellar core 
L1689B (see color scale on the right in MJy/sr). 
The 1.3~mm continuum emission map obtained by André et al. (1996) at the IRAM 30~m telescope is superposed as green contours 
(levels: 10, 30, 50~mJy/13\arcsec beam).
(b) Column density profile of L1689B (crosses)
%in the North-South direction (crosses)
derived from the absorption map shown in (a) by averaging the intensity over 
elliptical annuli for a 40$^\circ $ sector in the southern part of the core. 
The dashed curves show the most extreme profiles compatible with the data 
given the uncertainties 
%on the background and foreground 
affecting the absorption analysis.
The blue, solid curve represents the best fit of a Bonnor-Ebert sphere
model (embedded in a medium of uniform column density), 
obtained with the following parameters:
$\rho_c/\rho_{out} = 40 \pm 15$ (i.e., well into the unstable regime), 
$T_{eff} = 50 \pm 20$~K, 
$P_{ext}/k_B = 5\pm 3 \times 10^5\, \rm{K}\, \rm{cm}^{-3} $.
For comparison, the dotted line shows the $N\h2 \propto \bar{r}^{-1}$ 
profile of a SIS at $T = 10$~K.  
(Adapted from Bacmann et al. 2000.) 
\label{fig-abs}}
\end{figure}

%\noindent
More recently, it has been possible to use the {\it absorption} approach, 
both in the mid-IR from space (e.g. Bacmann et al. 2000, 
Siebenmorgen \& Kr\"ugel 2000) and in the near-IR from the ground 
%around 2~$\mu$m with ground-based near-IR arrays 
(e.g. Alves et al. 2001).
In particular, using the ISOCAM infrared camera aboard the $ISO$ satellite, 
Bacmann et al. (2000) carried out a 7~$\mu$m survey of 24 low-mass 
starless cores (all undetected by $IRAS$) 
and observed absorption features in 23 of them. As an example, 
Fig.~\ref{fig-abs}a shows the ISOCAM image  obtained 
at 6.75~$\mu$m for the prestellar core L1689B in the Ophiuchus 
complex ($d \sim 150$~pc). The core is seen in absorption 
against the diffuse mid-IR background arising from 
the rear side of the parent molecular cloud (e.g. Bernard et al. 1993, 
see also \S ~2.3 below). 
%Note the good correspondence between 
%the infrared absorption seen in the ISOCAM image and the 
%1.3~mm emission mapped with the IRAM~30~m telescope 
%(shown as contours in Fig.~\ref{fig-abs}a), confirming that both trace 
%the same cold dust material.

The column density structure of the absorbing cores can be derived 
from a simple modeling of the ISOCAM observations.
The mid-IR intensity measured at projected radius $\bar{r}$ from core center
%across the images 
may be expressed as:
\begin{equation}
        I(\bar{r}) = I_{back} \cdot e^{-\tau(\bar{r})} + I_{fore},  
\end{equation}
where $\tau $ is the core optical depth (directly related to the column density
N$_{H_{2}}$ via the dust opacity) and $I_{back}$ and $I_{fore}$ are the
%(uncertain) 
background and foreground intensities, respectively.  
Both $I_{back}$ and $I_{fore}$ are uncertain but their values are well 
constrained when independent millimeter measurements of the H$_{2}$ 
column density in the centers and outer parts of the cores are available
(see Bacmann et al. 2000 for details).
%Good constraints on $I_{back}$ and $I_{fore}$ can be obtained from 
%independent millimeter measurements of the H$_{2}$ column density 
%in the cores:  Millimeter continuum maps provide good estimates of 
%the central column densities, N$_{cen}$, accurate to a factor of 
%$\sim 2$ (see \S ~2.2 below); 
%C$^{18}$O(1-0) line observations provide
%estimates (also accurate to a factor of $\sim 2$) of the 
%column density N$_{out}$ of the cores at large radii.

Using these constraints on $I_{back}$ and $I_{fore}$, 
%(or alternatively N$_{peak}$ and N$_{out}$), 
a range of possible column density profiles can be derived for each core 
imaged in absorption with ISOCAM. This is shown in Fig.~\ref{fig-abs}b 
for L1689B, where the crosses display our `best' estimate of 
the column density profile (derived using a revised estimate of 
$8 \times 10^{22}$~cm$^{-2}$ for the central column density from 
millimeter continuum data, in agreement with a central temperature of 
$\sim 8$~K -- see \S ~2.3 below -- instead of 12.5~K as assumed by  
Bacmann et al. 2000).  
The dashed curves lying above and below the `best' profile show the most 
extreme column density profiles that are consistent with the 
mid-IR absorption 
data, given the range of permitted values for $I_{back}$ and $I_{fore}$. 
In general, four different regimes can be distinguished on the 
prestellar column density profiles 
%derived by Bacmann et al. (2000)
%from their ISOCAM absorption data 
(see, e.g., Fig.~\ref{fig-abs}b):

a) a flat inner region (of radius $R_{flat} = 5000 \pm 1000$~AU for L1689B 
according to the fitting analysis shown in Fig.~\ref{fig-abs}b), 
b) a region roughly consistent with $N\h2 \propto \bar{r}^{-1}$ 
(corresponding 
to $\rho \propto r^{-2}$ for a spheroidal core), 
c) an edge where the column density falls off typically more rapidly than 
$N\h2 \propto \bar{r}^{-2}$ with projected radius 
(suggesting a density gradient steeper than $\rho \propto r^{-3}$),  
until d) the end of the core is reached 
(at $R_{out} = 28000 \pm 1000$~AU for L1689B) 
and $N\h2 $ fluctuates about the 
mean value $N_{out} \sim 1-2 \times 10^{21}$~cm$^{-2}$ characterizing 
the ambient molecular cloud.

A qualitatively similar column density profile was derived by 
Alves et al. (2001) for the Bok globule B68 (for which 
$R_{flat} \approx 4000$~AU and $R_{out} \approx 12500$~AU), based on VLT 
and NTT measurements of the near-IR colors of background stars seen through, 
and suffering extinction from, this globule. 

Thus, the {\it inner flattening of the density profiles of prestellar cores}
first observed in submillimeter emission (see above) is confirmed by 
infrared absorption/extinction measurements, which are essentially 
{\it independent of the temperature distribution}. 
Moreover, a new result emerges: at least in some cases, such as L1689B 
and B68, isolated prestellar cores 
appear to be characterized by {\it sharp edges} defining typical outer radii 
$R_{out} \sim 20000\, \rm{AU} \sim 0.1\, \rm{pc} $.

\subsection{Comparison with theoretical models}

These observed features, i.e., flat inner region and 
sharp outer edge, set constraints on possible models 
of core structure.   
First of all, it is clear that self-similar, singular models such as  
the SIS or magnetized, 
non-spherical generalizations of it (e.g. Shu 1977, Li \& Shu 1996), 
cannot account for the detailed 
density structure of prestellar cores.
As pointed out by Bacmann et al. (2000), the logotropic 
models of McLaughlin \& Pudritz (1996), which behave as $\rho \propto r^{-1}$
over a wide range of radii, are also inconsistent with the observations of
(low-mass) starless cores.
On the other hand, the circularly-averaged column density profiles 
can often be fitted remarkably well with models of pressure-bounded 
Bonnor-Ebert spheres, as first demonstrated by Alves et al.~(2001) for B68. 
This is also the case of L1689B as illustrated in Fig.~\ref{fig-abs}b. 
The quality of such fits shows that 
Bonnor-Ebert spheres provide a good, first order model for the structure of
isolated prestellar cores. In detail, however, there are several problems with
these Bonnor-Ebert models. First, most starless cores 
exhibit elongated shapes (see, e.g., Fig.~\ref{fig-intro}a and
Fig.~\ref{fig-abs}a) 
which are not consistent with spherical models beyond $\sim 0.03$~pc. 
Second, the inferred density contrasts (from center to edge) 
are generally larger (i.e., $\simgt $~20--80 -- see Fig.~\ref{fig-abs}b) 
than the maximum contrast of $\sim $~14 for a stable Bonnor-Ebert 
sphere (cf. \S ~1.1).
Third, the effective core temperature needed in these models 
(for thermal pressure gradients to balance self-gravity)
is often significantly larger than both the average dust temperatures 
measured 
with ISOPHOT (e.g. Ward-Thompson et al. 2002) and the gas temperatures measured
in NH$_3$ (e.g. Lai et al., this volume). In the case of 
L1689B, for instance, the effective temperature of the Bonnor-Ebert fit 
shown in Fig.~\ref{fig-abs}b is $T_{eff} \sim 50$~K, while the dust temperature
observed with ISOPHOT is only $T_d \sim 11$~K (Ward-Thompson et al. 2002).
Finally, the physical process responsible for bounding the cores at some
external pressure is unclear. These arguments suggest that prestellar 
cores cannot be described as simple isothermal 
hydrostatic structures and are either already contracting (see \S ~3.1 below)
or experiencing extra support from static or turbulent magnetic fields 
(e.g. Curry \& McKee 2000).

One way of accounting for large density contrasts and high effective
temperatures is to consider models of cores threaded by 
a static magnetic field and evolving through ambipolar diffusion 
(e.g. Ciolek \& Mouschovias 1994, Basu \& Mouschovias 1995). 
In these models, at any given time prior to protostar formation, 
the cores are expected to feature a 
uniform-density central region whose size corresponds to the instantaneous 
Jeans length. This agrees well with the characteristics of the flat inner 
regions seen in starless cores. Furthermore, the observed 
sharp edges (e.g. Fig.~\ref{fig-abs}b) are consistent with the model 
predictions (shortly) after the formation of a 
magnetically supercritical core. 
Physically, this is because when a supercritical core 
forms, it collapses dynamically inward, while the outer, subcritical 
envelope is still efficiently supported 
by the magnetic field and remains essentially ``held in place''. As a result, 
a steep density profile develops at the outer boundary of the 
supercritical core (see Fig.~8 of Basu \& Mouschovias 1995).\\
A problem, however, with these ambipolar diffusion models involving 
only a static magnetic field is that they require fairly large  
field strengths ($\sim $~30--100~$ \mu$G, see Bacmann et al. 2000), 
which seem to exceed the (few) existing Zeeman measurements 
for low-mass dense cores (e.g. Crutcher 1999, Crutcher \& Troland 2000).
It is possible that more elaborate versions of the models, 
incorporating the effects of a non-static, turbulent magnetic field 
in the outer parts of the cores and in the ambient cloud, 
would be more satisfactory and could 
also account for the filamentary shapes often seen on large 
($\simgt 0.25$~pc) scales (cf. Curry 2000, 2002 and Jones \& Basu 2002).

\subsection{Temperature distribution in starless cores}

\begin{figure}[ht!]
\vspace{-0.5cm}
\centering
\includegraphics[width=0.85\linewidth]{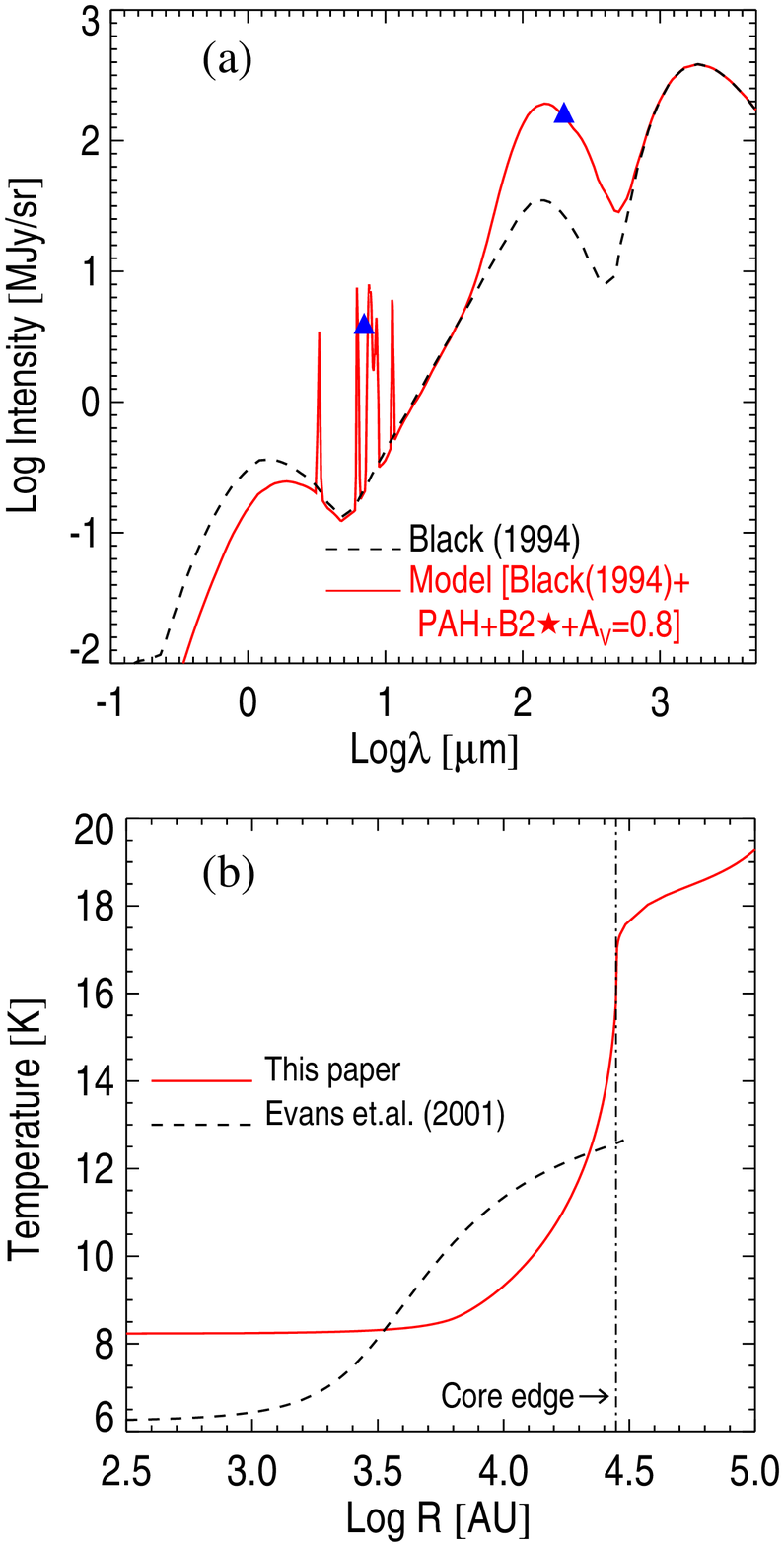}
\caption{(a) Effective radiation field impinging on L1689B 
(red, solid curve), as estimated from 
the observed levels of the mid-IR and far-IR backgrounds 
(blue triangles, see text). This is much stronger 
than the Black (1994) radiation field (black, dashed curve).
(b) Model dust temperature profiles calculated for L1689B with MODUST 
under two sets of assumptions: the red, solid  
curve assumes the effective radiation field shown in (a), the 
density profile derived by Bacmann et al. (2000) 
(with $R_{flat} = 6600$~AU), and the dust opacity law of 
Preibisch et al. (1993) with thin ice mantles in the core and no ice 
mantles in the ambient cloud; the black, dashed curve corresponds to
the model favored by Evans et al.~(2001) which has $R_{flat} = 2000$~AU 
(for $d = 160$~pc), no ambient cloud, and assumes the Black (1994) 
radiation field $\times 0.5$ 
and the dust opacity law of Ossenkopf \& Henning (1994) with 
%thin ice mantles and 
coagulation (OH5).
%for two Bonnor-Ebert-like density profiles 
%with $R_{flat} = 6600$~AU (cf. Bacmann et al. 2000) and 
%$R_{flat} = 3300$~AU (smallest value consistent with the 1.3~mm constraints),
%respectively. 
\label{isrf_temp}}
\end{figure}

The models discussed in \S ~2.2 above assume that prestellar cores are 
isothermal, which is quite a good first approximation (e.g. Larson 1969, 
Tohline 1982). In actual fact, however, there are good reasons to believe
that the central regions of starless cores are somewhat cooler than their 
outer regions. Indeed, the results of $ISO$ imaging in the mid-IR
(Bacmann et al. 2000) and far-IR (Ward-Thompson et al. 2002) 
are consistent with the idea that starless cores 
are heated only from outside by the local interstellar radiation field (ISRF): 
the amount of energy radiated by such cores in the far-IR is roughly equal 
to the fraction of ISRF energy absorbed at shorter (e.g. mid-IR) wavelengths, 
with no evidence for any central heating source 
(Ward-Thompson et al. 2002). In such a situation, recent dust radiative 
transfer calculations (Evans et al. 2001; Zucconi et al. 2001)
%of the thermal energy balance
predict that there should be a positive temperature gradient from the cores'
centers (with $T_d$ as low as $\sim $~5--7~K) to their edges (at 
$T_d \sim $~15~K). [Note that, in the dense 
($\nhd > 3 \times 10^4 \rm{cm}^{-3}$) inner regions, one expects the  
temperature of the gas to be well coupled to that of the dust 
(e.g. Doty \& Neufeld 1997).] The first calculations published by 
\cite{Evans01} and \cite{Zuc01} assumed rather a simplistic input radiation
field, based on current estimates of the average radiation field in the 
solar neighborhood (e.g. Black 1994). We have more recently carried out 
similar calculations in which we use the levels of the diffuse mid-IR 
and far-IR backgrounds observed toward the cores
to make more realistic estimates of the effective radiation field
directly impinging on their surfaces. 
Fig.~\ref{isrf_temp}a gives an illustration 
for the prestellar core L1689B already discussed in \S ~2.1  
(see Fig.\ref{fig-abs}). In this case,
the radiation field differs (in strength and spectrum) from the Black 
standard radiation field for several reasons: 
(1) due to the presence of early-type stars
such as HD~147889 (B2~V) only a few parsecs away
%$\sim 6-10$~pc 
from the Ophiuchus cloud (e.g. Liseau et al. 1999), the ambient far-UV field 
is significantly stronger (i.e., $G_0 \sim 12$) than the solar neighborhood
average; (2) this strong FUV field in turn excites small PAH-like grains in 
the outer ($A_V \simlt 1$) layers of the cloud which are responsible for a 
high level of diffuse mid-IR emission distributed in several spectral bands 
(cf. Bernard et al. 1993 and Fig.~\ref{isrf_temp}a); 
(3) the L1689B core is itself embedded at some depth ($A_V \sim 1$) 
inside the cloud, 
so that a good fraction of the external FUV radiation is effectively 
reprocessed to longer wavelengths. Consequently, the effective radiation field
for L1689B is about {\it one order of magnitude stronger} 
in both the mid-IR and the
far-IR than the Black radiation field (cf. Fig.~\ref{isrf_temp}a), 
in agreement with the strong mid-IR and far-IR backgrounds measured by 
Bacmann et al. (2000) and Ward-Thompson et al. (2002) with ISOCAM and ISOPHOT 
($\sim 4$~MJy/sr at 7~$\mu$m and $\sim 120$~MJy/sr at 200~$\mu$m, 
as opposed to $\sim 0.25$~MJy/sr and $\sim 30$~MJy/sr, respectively,  
in the Black radiation field). 
Based on this estimate of
the effective radiation field, we have used the MODUST radiative transfer code 
(e.g. Bouwman 2001, Kemper et al. 2001) to compute the temperature 
distribution in  
the L1689B core for several plausible density distributions consistent with 
the 1.3~mm continuum emission map of Andr\'e et al. (1996) and the 7~$\mu$m 
absorption map of Bacmann et al. (2000). The results are illustrated in
Fig.~\ref{isrf_temp}b.

\begin{figure}[ht!]
\vspace{-0.2cm}
%\resizebox{\hsize}{!}{\includegraphics{waterloo_struc.ps}}
\resizebox{\hsize}{!}{\includegraphics{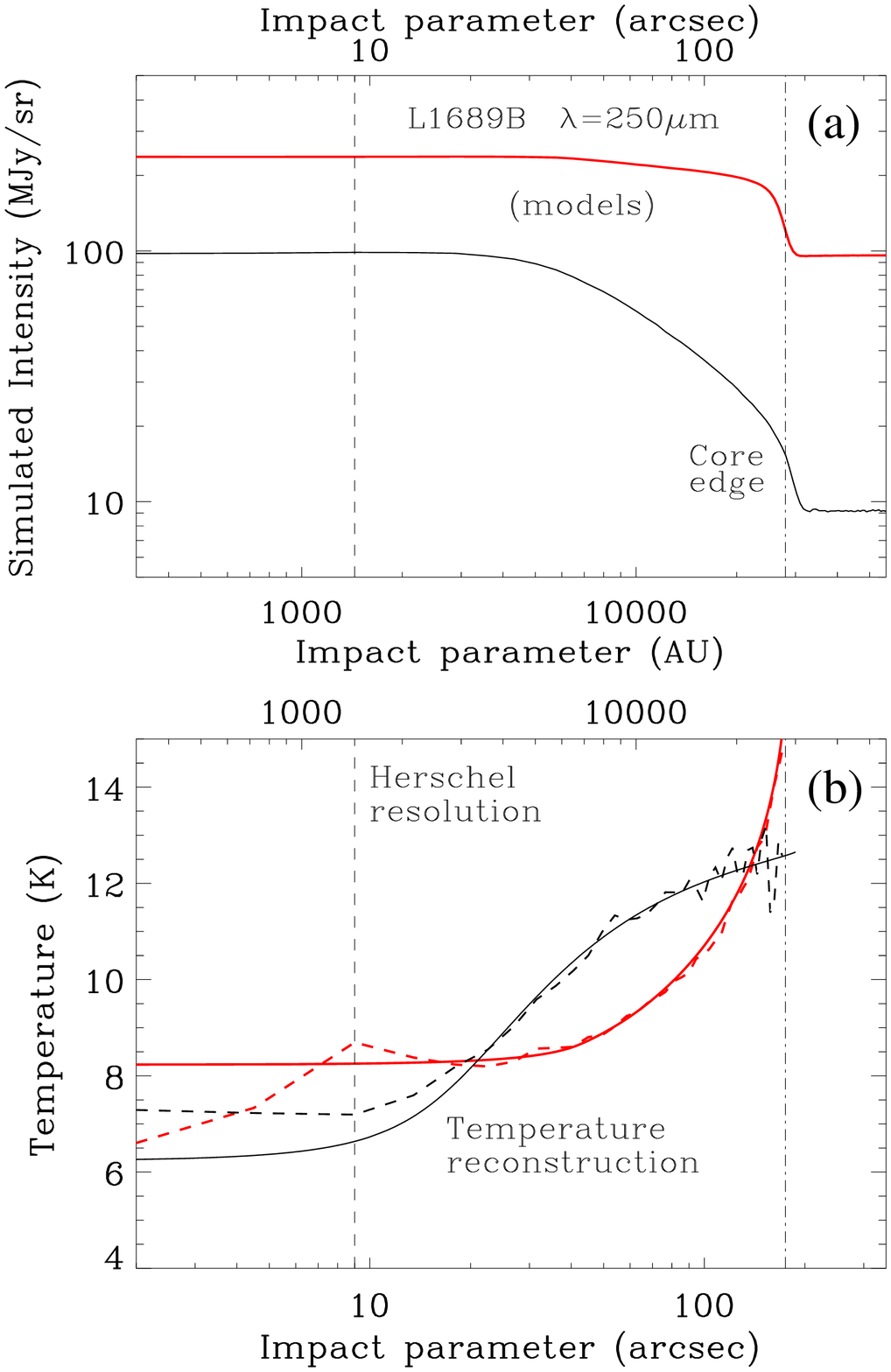}}
\caption{Simulated $Herschel$ observations of the radial structure of the 
prestellar core L1689B. The top panel (a) shows two model intensity profiles,
consistent with present millimeter data, as they could be observed
by $Herschel$ at 250~$\mu $m. The upper (red) curve corresponds to the model we 
favor here based on our best estimate of the effective radiation field 
(see Fig.~\ref{isrf_temp}), while the lower curve is for the model proposed
by Evans et al.~(2001) based on the Black (1994) ISRF $\times 0.5$. 
Note the large differences in shape and absolute scaling between the two models.
The bottom panel (b) shows the dust temperature distributions
(solid curves) calculated with MODUST for these two models.
The dashed curves (almost indistinguishable from the solid 
curves beyond 20\arcsec) show the reconstructed temperature profiles  
that could be derived from $Herschel$ maps in 6 bands between 
75~$\mu$m and 500~$\mu$m (plus ground-based data at 850~$\mu$m and 1.3~mm), 
assuming the 3-D core geometry is known. The reconstruction becomes uncertain 
below the $Herschel$ resolution limit (9\arcsec HP beam radius at 250~$\mu$m).
\label{herschel}}
\end{figure}

\begin{figure*}
\centering
\vspace{-2.5cm}
\includegraphics[width=0.6\linewidth,angle=270]{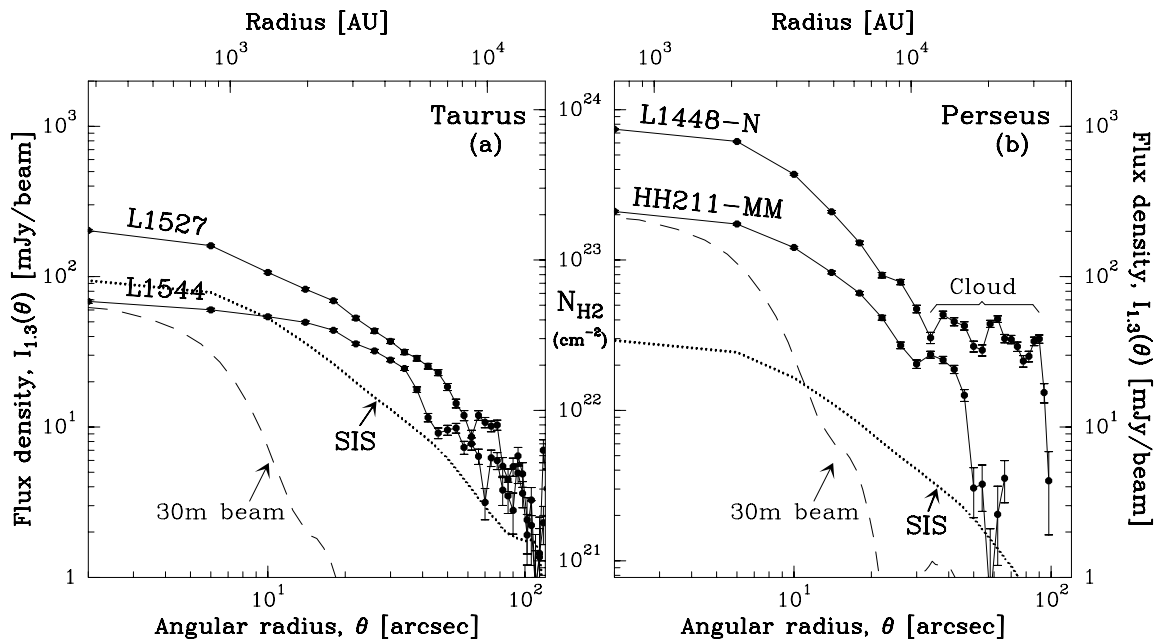}
\caption{Averaged 1.3~mm radial intensity profiles of (a) L1544 
(prestellar) and L1527 (Class~0) in Taurus ($d = 140$~pc), 
and (b) HH211--MM and L1448-N (Class~0s) in Perseus ($d = 350$~pc), 
compared to a synthetic profile, simulated for a singular isothermal 
sphere (SIS) model at $T = 10 $~K (see Appendix of Motte \& Andr\'e 2001 
for details about such simulations).
%(dotted curve). 
The profile of the 30m beam is shown.
%as a dashed curve. 
An approximate column density scale, calculated assuming 
representative dust properties ($T_d = 15$~K, $\kappa_{1.3} = 0.0075\ \cmg$) 
is also indicated. (Adapted from Ward-Thompson et al. 1999 and 
Motte \& Andr\'e 2001.) 
\label{fig-prof}}
\end{figure*}

Our calculations confirm some of the conclusions of \cite{Evans01} and
\cite{Zuc01}, namely that the temperature reaches a minimum $< 10$~K in the 
centers of prestellar cores and that the central temperature depends 
primarily on the central optical depth (directly related to the 
degree of shielding from the external ISRF). However, we find that  
the minimum temperature may not be as low as reported by \cite{Evans01} 
and \cite{Zuc01} and that the shape of the temperature profile may be 
quite different: it can be seen in Fig.~\ref{isrf_temp}b that the major drop 
in temperature occurs in the outer parts of the core in our preferred model 
of L1689B (red curve), while it occurs closer to the center in the 
\cite{Evans01} model (dashed curve).  
We also stress that all of the models we have run have a relatively 
high mass-averaged 
dust temperature of $\sim 11$~K, comparable to the temperature estimated by 
fitting a greybody to the global SED (cf. Fig.~13 of 
Ward-Thompson et al. 2002). (Note that Doty \& Palotti 2002 have independently 
reached a similar conclusion.) 
Furthermore, if our
assumptions about the effective radiation field and the dust opacity law 
(see Fig.~\ref{isrf_temp}b) are correct, then the radius of the flat inner 
region in the density profile of L1689B must lie between 
$R_{flat} = 3300$~AU (smallest value consistent with the 1.3~mm constraints
of Andr\'e et al. 1996) and $R_{flat} = 6600$~AU (value found by 
Bacmann et al. 2000 assuming a central core temperature of 12.5~K -- see \S ~2.1).
A value as small as $R_{flat} = 2000$~AU (cf. Evans et al. 2001) is ruled out 
according to our analysis. Thus, conclusions derived from 
(sub)millimeter emission maps on the density structure of prestellar cores 
under the simplifying hypothesis that the dust temperature is uniform 
(e.g. Ward-Thompson et al. 1999) are probably not seriously flawed.

In our opinion, 
the final word on the structure of 
prestellar cores will come from high-resolution mapping at far-IR and
submillimeter wavelengths with future space-borne telescopes such as 
$Herschel$ to be launched by ESA in 2007 (e.g. Pilbratt et al. 2001).
Using $Herschel$ images at 75--500~$\mu$m (in combination with
ground-based dust continuum mapping at longer submillimeter wavelengths)
to construct SED {\it maps} for at least the nearest, spatially resolved
prestellar cores, it will be possible to reconstruct their 
{\it intrinsic temperature and density distributions simultaneously} 
as illustrated in Fig.~\ref{herschel}. Such multi-band mapping observations
will also set strong constraints on 
the {\it dust emissivity properties} (e.g. the opacity index $\beta $) 
and their spatial variation across dense cores.

\subsection{Structure of protostellar envelopes}

In the case of {\it protostellar} envelopes, the accretion luminosity released 
close to the central protostar provides an additional source of heating 
compared to starless cores. In practice, however, the thermal structure 
of the envelopes surrounding low-luminosity protostars 
($L_{bol} \simlt $~1--10~$L_\odot$), including most Taurus embedded YSOs,
is strongly affected by external heating from the ambient ISRF 
(Shirley et al. 2002), and the simple isothermal assumption 
(with T$\,\simeq 10\ $K) is often not too bad over a wide range of radii 
($r \sim $~3000--15000~AU -- cf. Motte \& Andr\'e 2001).

The density structure of YSO envelopes has been studied primarily
with the (sub)millimeter emission approach (although see Harvey et al. 2001
for a recent extinction study).
In contrast to prestellar cores, protostellar envelopes are always 
found to be strongly centrally condensed (see, e.g., Fig.~\ref{fig-intro}b) 
and do {\it not} exhibit any marked inner flattening in their 
radial (column) density profiles.   
This is illustrated in Fig.~\ref{fig-prof} which 
shows the circularly-averaged radial intensity profiles measured at 1.3~mm 
for the circumstellar envelopes of the Class~0 objects L1527 in Taurus and 
HH211-MM and L1448-N in Perseus (Motte \& André 2001). (Note that 
the inner disk contribution has been subtracted from the profiles based 
on high-resolution interferometric measurements with the 
IRAM Plateau de Bure interferometer.)
%-- Motte 1998).
Apart from a small overdensity factor ($\sim 2$), the profile of the 
L1527 envelope closely follows the {\it shape} 
of the SIS model profile 
(see Fig.~\ref{fig-prof}a), and 
is significantly steeper than the profile of the prestellar core L1544 
at small radii.

More generally, several (sub)millimeter continuum stu-dies 
indicate that protostellar envelopes 
in regions of isolated star formation such as Taurus 
have radial density gradients consistent with 
$\rho(r) \propto r^{-p}$ with $p \sim $~1.5--2 
over more than $\sim $~10000--15000~AU in radius 
(e.g. Ladd et al. 1991, Chandler \& Richer 2000, Hogerheijde \& Sandell 2000,
Shirley et al. 2000, Motte \& André 2001, J\o rgensen et al. 2002).
% 
%Notice that, to first order, the envelopes of low-luminosity protostars 
%($L_{bol} \simlt $~1--10~$L_\odot$), including most Taurus embedded YSOs, 
%should be roughly  
%isothermal (at the temperature T$\,\simeq 10\ $K of the parent cloud) 
%over the range of radii ($r \sim $~1000--15000~AU) probed by IRAM~30~m maps. 
%
The density gradient estimated for (Class~0 and Class~I) protostars
(i.e., $p \sim $~1.5--2) agrees with most  
collapse models which predict a value of $p$ between 2 and 1.5  
during the protostellar accretion phase (before and after the passage of 
the collapse expansion wave, respectively -- see \S ~1.1 above).  
Furthermore, the median envelope mass    
measured by Motte \& Andr\'e~(2001) for the bona-fide protostars of Taurus 
[$\overline{M_{env}} (r < 4200 AU) \sim 0.3\, M_\odot $] is roughly consistent  
with the predictions of the inside-out collapse model of Shu~(1977) 
for $\sim 10^5$~yr-old protostars.

The situation is markedly different in regions where stars form in 
tight groups or {\it clusters}, such as Serpens, Perseus, and the 
$\rho$~Oph main cloud. In this case, the observed envelopes 
are clearly {\it not scale-free}: 
they merge with dense cores, other envelopes, 
and/or the diffuse ambient cloud at a {\it finite} 
radius $\rout \simlt 5\,000$~AU (Motte, Andr\'e, \& Neri 1998;  
Motte \& Andr\'e 2001; see also Fig.~5 of Mundy, Looney, \& Welch 2000). 
Moreover, as can be seen in Fig.~\ref{fig-prof}b for HH211-MM and L1448-N, 
the Class~0 envelopes mapped in cluster-forming regions  
are found to be {\it 3 to 12 times denser} than the SIS at T$\,=10\ $K 
(which is the typical gas kinetic temperature expected in these clouds prior 
to massive star formation -- see Goldsmith \& Langer 1978, Evans 1999, 
Goldsmith 2001, and Fig.~\ref{isrf_temp}b).  
Static magnetic fields can only account for moderate overdensity 
factors ($\simlt 2$) compared to an unmagnetized SIS (see Li \& Shu 1997). 
Turbulence could in principle contribute to the support of the initial
dense core (e.g. Myers \& Fuller 1992, Mardones et al. 1997).  
However, the small-scale condensations corresponding to the
precursors of protostars in star-forming clusters (e.g. Motte et al. 1998) 
appear to be largely devoid of turbulence (see \S ~3.1 below).
%in agreement with the `kernel' picture (Myers 1998). 
A more likely explanation is that, in such regions, protostellar collapse 
is {\it induced by strong external disturbances} and 
starts from {\it non-singular initial conditions}, resulting in a 
{\it non-equilibrium 
density configuration} with a large overdensity factor ($\simlt 10$) and 
significant inward velocities near $t \sim 0 $ (see also \S ~4 and 
Fig.~\ref{hennebelle} below).
%point-mass formation
%(as in the Larson-Penston similarity collapse flow, see \S ~1.1 above). 
%Follow-up line observations of the envelopes' velocity fields are 
%nevertheless required to confirm this suggestion (see \S ~3.2 below).

\begin{figure}[ht!]
%\resizebox{\hsize}{!}{\includegraphics[angle=270]{posvit_mix_mod_wloo.eps}}
\resizebox{\hsize}{!}{\includegraphics[angle=270]{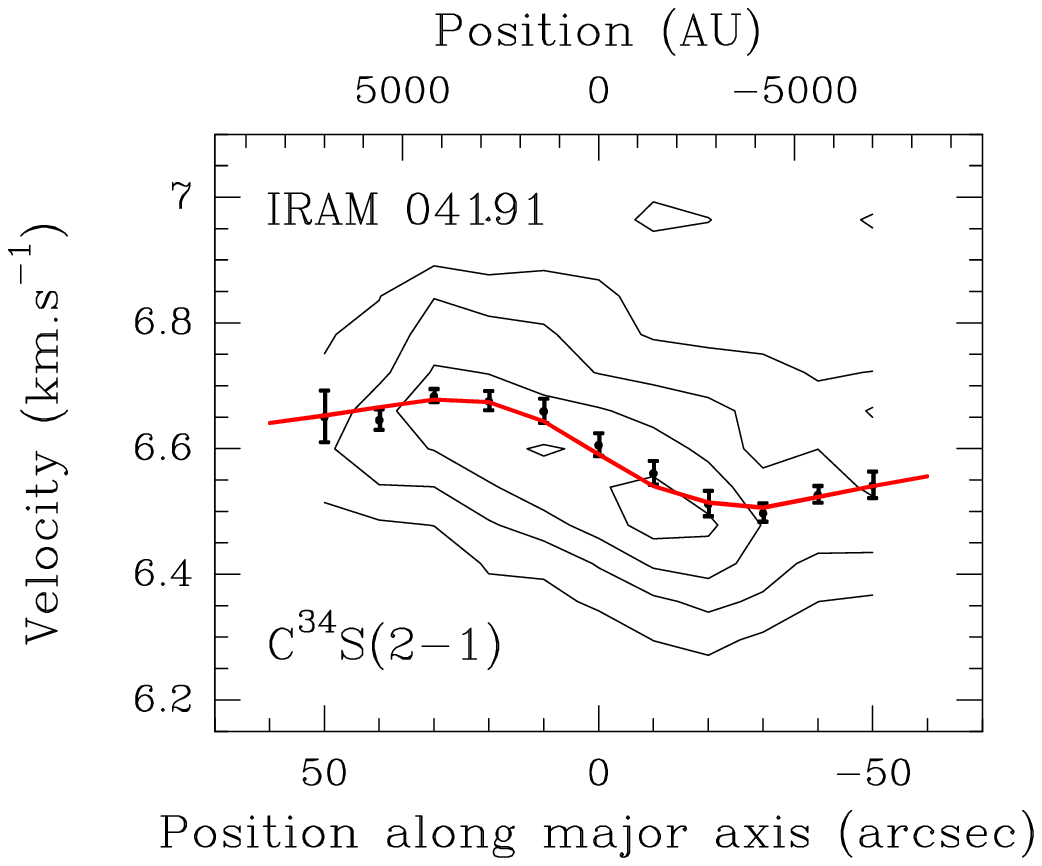}}
\caption{Position-velocity diagram along the major axis of the  
IRAM~04191 envelope (i.e., perpendicular to the flow) 
based on a C$^{34}$S$(2-1)$ 
%C$_3$H$_2$(2--1) on-the-fly 
map taken at the IRAM 30m telescope (Belloche et al. 2002).
Contours: 0.2 to 0.8 by 0.2~K.
%0.3 to 1.8 by 0.3~K.
The dots with error bars mark the observed velocity centroids. 
%of the underlying spectra are marked by crosses. 
The solid curve shows the profile of a model with differential 
rotation ($V_{rot}(r >3500\, \rm{AU} ) \propto r^{-1.5}$) 
also used in Fig.~\ref{infall} (see text). 
\label{rotation}}
\end{figure}

\begin{figure*}
\centering
\includegraphics[width=0.8\linewidth]{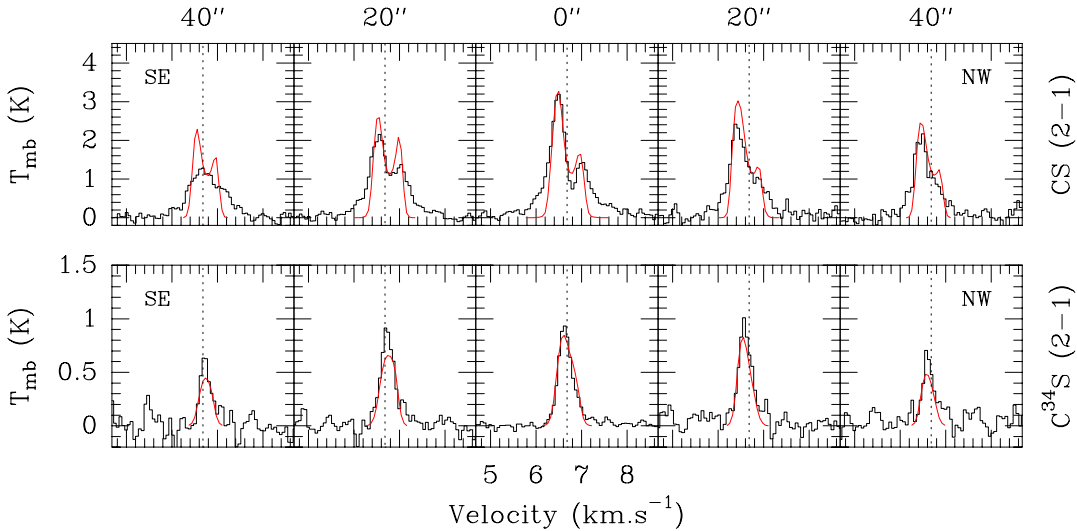}
\caption{Spectra observed with the IRAM 30m telescope toward IRAM~04191, along 
the axis perpendicular to the outflow, in the optically thick 
CS$(2-1)$ and optically thin C$^{34}$S$(2-1)$ lines.
Synthetic spectra corresponding to the `best-fit' collapse model 
including rotation found by Belloche et al. (2002)
%with $\rho \propto r^{-1.5}$, 
%a constant infall velocity $V_{inf} = 0.2\, \kms $ up to $R_{inf} = 10000$~AU.
%$V_{inf} \propto r^{-0.5}$, and 
%$V_{inf}$(700~AU)~$ = 0.4\, \kms $, $V_{inf}$(2000~AU)~$ = 0.1\, \kms $, 
%and $V_{rot}$(2000~AU)~$ = 0.2\, \kms $ 
are superposed (see text for parameters).
\label{infall}}
\end{figure*}

\section{Velocity Structure}

\subsection{Internal motions in starless cores}

On large ($\simgt $~pc) scales, the spectral line profiles/widths observed 
in molecular clouds are indicative of supersonic turbulent 
motions (e.g. Falgarone \& Phillips 1990). 
The nonthermal component of the line width, $\sigma_{NT} $, roughly follows the
scaling law $\sigma_{NT} \propto R^{0.5} $ (e.g. Fuller \& Myers 1992), which  
is believed to arise from MHD turbulence (e.g. Mouschovias \& Psaltis 1995). 
On intermediate ($\simgt 0.1$~pc) scales, 
there are strong differences between the ``low-mass'' cores observed in 
regions of distributed star formation (e.g. Taurus), whose line widths are 
dominated by thermal motions (e.g. Benson \& Myers 1989), and the 
``massive'', cluster-forming cores of, e.g., Ophiuchus and Orion, whose
line widths are still dominated by turbulent motions 
(e.g. Caselli \& Myers 1995). 
On small ($\simlt 0.03$~pc) scales, however, the   
prestellar condensations of the Ophiuchus, Serpens, Perseus, and Orion  
cluster-forming regions are always characterized by very {\it narrow line widths}
(e.g. Belloche et al. 2001, Myers 2001), which is reminiscent 
of the thermal cores of Taurus. 
In the $\rho$~Oph protocluster, for instance, the nonthermal 
velocity dispersion is about half the thermal velocity dispersion of H$_2$ 
($\sigma_{NT}/\sigma_{T} \sim 0.7 $) toward the starless condensations of
the dense cores Oph~B1, C, E, F (Belloche et al. 2001). This indicates  
that, {\it even in cluster-forming clouds}, the initial conditions for
individual protostellar collapse are ``coherent'' (cf. Goodman et al. 1998) 
and essentially free of turbulence.
The dissipation, on small ($< 0.1$~pc) scales, of a significant fraction of 
the turbulent motions observed on large scales is thus a prerequesite for 
the formation of prestellar condensations 
(e.g. Nakano 1998, Myers 1998, Falgarone \& Pety 2001). 
Apart from low levels of turbulence, low-mass prestellar cores are also
characterized by small rotation rates in general 
($\Omega \sim 10^{-14} - 10^{-13}$~rad~s$^{-1}$ --
e.g. Goodman et al. 1993), and subsonic, extended infall motions 
(e.g. Lee, Myers, \& Tafalla 2001 -- see also Evans, this volume).

\subsection{Rotation and infall in Class~0 objects}

The velocity structure of only a few Class~0 objects has been studied 
in detail up to now. Here, we mostly discuss the example of 
the nearby ($d = 140$~pc) and relatively isolated 
Class~0 object IRAM~04191$+$1522 (IRAM~04191 for short) 
found by 
Andr\'e, Motte, \& Bacmann (1999) in the vicinity of the Taurus 
Class~I source IRAS~04191$+$1523 (e.g. Tamura et al. 1991 -- 
see Fig.~\ref{fig-intro}b).
 
IRAM~04191 features a very high envelope mass to luminosity ratio 
($M^{< 4200AU} _{env}/L_{bol} \sim 3\ M_\odot/L_\odot $)
and a low bolometric temperature ($T_{bol} \sim 18$~K), suggesting an age $t \simlt 3 \times 10^4$~yr since the beginning of the accretion phase 
(see Andr\'e et al. 1999).
It is also remarkable in that it has a very low bolometric luminosity 
($L_{bol} \sim 0.15\, L_\odot $), and no or only a tiny accretion disk 
($R_{disk} < 15$~AU, $M_{disk} < 2 \times 10^{-3}\, M_\odot$). 
%and features a jet-like outflow (with $V_{flow} \sim 10\, \kms $ 
%and $\fcom \sim 1.5 \times 10^{-5} \mkmsyr $ -- cf. Fig.~\ref{fig-intro}b), 
%as well as extended infall motions (e.g. Fig.~\ref{infall}, see below). 
Accordingly, {\it IRAM~04191 is probably the youngest accreting 
protostar currently known in Taurus}. 

\begin{figure}[ht]
%\resizebox{\hsize}{!}{\includegraphics[angle=270]{vel_om_waterloo.eps}}
\resizebox{\hsize}{!}{\includegraphics[angle=270]{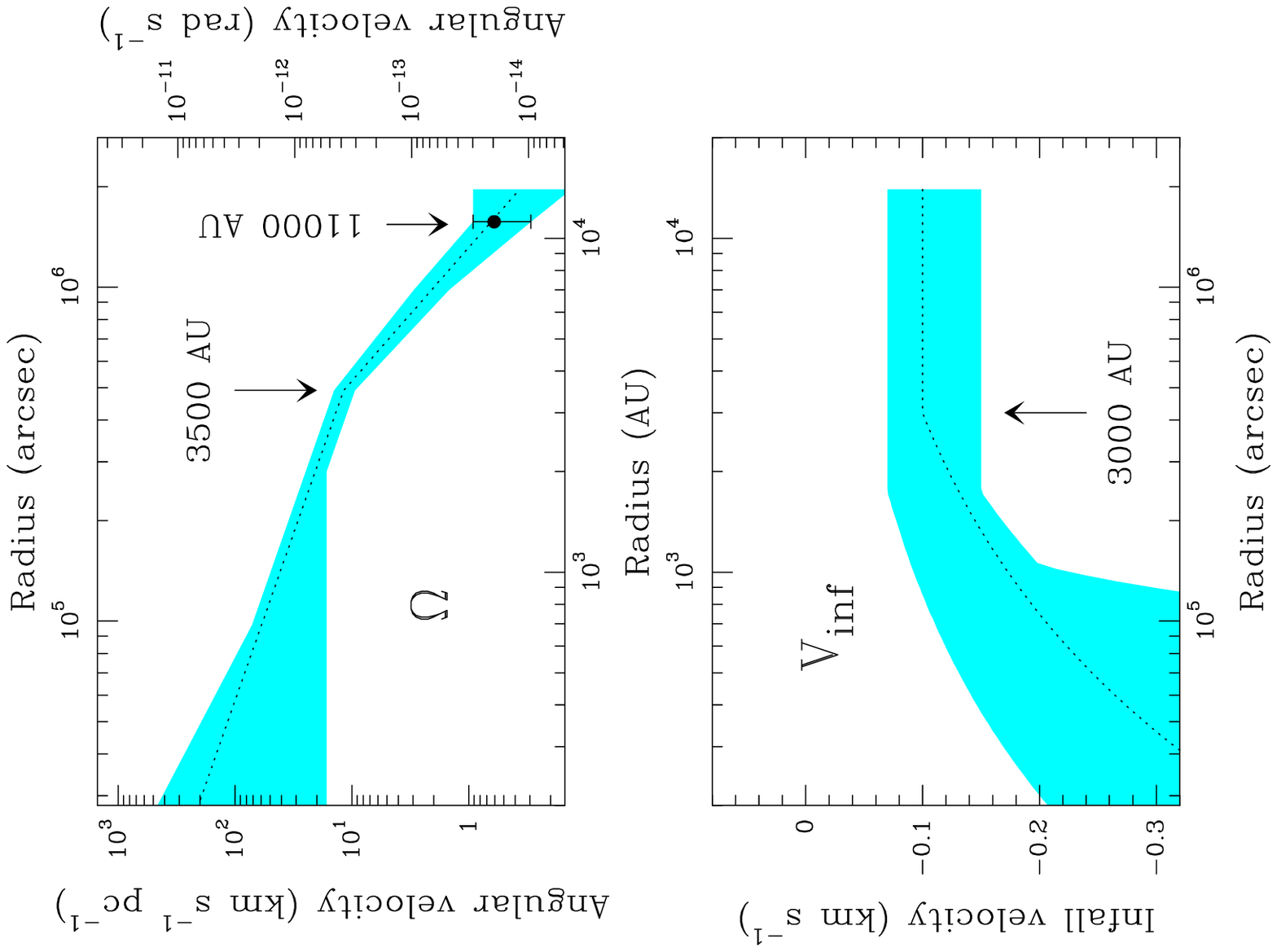}}
\caption{Rotational angular velocity (a) and infall velocity (b) 
inferred in the IRAM~04191 envelope based on the radiative transfer modeling 
of Belloche et al. (2002).
The shaded areas show the estimated domains where the 
models match the CS and C$^{34}$S observations reasonably well. 
\label{iram04191_model}}
\end{figure}

IRAM~04191 is associated with a flattened circumstellar 
envelope, seen in the dust continuum (Fig.~\ref{fig-intro}b) and in
dense gas tracers such as N$_2$H$^+$, C$_3$H$_2$, H$^{13}$CO$^+$, 
and DCO$^+$ (Belloche et al. 2002). 
All of the maps taken in small optical depth lines show a clear velocity 
gradient across this envelope of $\sim 7\, \kms \, $pc$^{-1}$ increasing from 
west to east along the long axis, i.e., roughly perpendicular to 
the outflow axis (see Fig.~\ref{fig-intro}b and Fig.~\ref{rotation}).
Strongly suggestive of rotation, this velocity gradient corresponds to
%implies that the IRAM~04191 envelope is rotating a factor of 
%$\sim $~2--30 faster than typical low-mass NH$_3$ cores (cf. 
%Goodman et al. 1993), with 
a mean angular velocity $\Omega \sim 3 \times 10^{-13}$~rad~s$^{-1}$ in the
inner $\sim 3000$~AU region (after deprojection from a viewing angle 
$i = 50^\circ $).
Furthermore, this rotation does not occur in a rigid-body, 
but differential, fashion: the inner $\sim $~3500~AU-radius region seems 
to be rotating significantly faster than the outer parts of the core 
(see Fig.~\ref{rotation}).  
The ``S'' shape of the position-velocity diagram shown in Fig.~\ref{rotation} 
can be fitted with a rotation curve of the form 
$V_{rot}(r) \propto r^{0.1 \pm 0.4} $ for $ r < 3500 $~AU
and $V_{rot}(r) \propto r^{-1.5 \pm 0.5} $ for $ 3500 < r < 11000 $~AU, 
corresponding to $\Omega(r) \propto r^{-0.9 \pm 0.4} $ and 
$\Omega(r) \propto r^{-2.5\pm 0.5} $, respectively 
(see Fig.~\ref{iram04191_model}a and Belloche et al. 2002 for details).

Direct evidence for infall motions over a large portion of the 
IRAM~04191 envelope is also observed in optically thick lines such as 
CS(2--1), CS(3--2), 
%HCO$^+$(1--0), 
H$_2$CO($2_{12} - 1_{11}$), 
and H$_2$CO($3_{12} - 2_{11}$). 
These lines are double-peaked and skewed to the blue up to 
$\simgt $~40\arcsec ~from source center (see Fig.~\ref{infall}), which is 
indicative of infall motions up to a radius 
$R_{inf} \simgt 5000$~AU from center 
(cf. Evans 1999 and Evans, this volume). 
Radiative transfer modeling confirms this view, suggesting a relatively 
flat infall velocity profile ($V_{inf} \sim 0.1\, \kms $) for 
$3000 \simlt r \simlt 11000$~AU and larger infall velocities scaling 
as $V_{inf} \propto r^{-0.5}$ for $r \simlt 3000$~AU
%(with $V_{inf} \sim  0.25\, \kms $ at 700~AU) in a small inner region 
(see Fig.~\ref{iram04191_model}b and Belloche et al. 2002 for details).
The mass infall rate is estimated to be 
$\dot{M}_\mathrm{inf} \sim 2-3 \times a_s^3/G 
\sim 3 \times 10^{-6}$~M$_\odot$~yr$^{-1}$ 
(with $a_s \sim 0.15-0.2\, \kms $ for $T \sim 6-10$~K), 
roughly independent of radius.

Another Class~0 object whose kinematics has been quantified in detail 
recently is the IRAS~4 system in the NGC~1333 protocluster 
(Di Francesco et al. 2001). IRAS~4 is a triple system (A/B/C) on scales 
of $\sim 10000$~AU, whose primary component (A) itself breaks up into 
a $\sim 600$~AU binary (e.g. Looney et al. 2000). Using the IRAM Plateau 
de Bure interferometer, Di Francesco et al. (2001) observed inverse P-Cygni 
profiles in H$_2$CO($3_{12} - 2_{11}$) toward IRAS~4A and IRAS~4B, from 
which they derived very large mass infall rates of 
$\sim 1.1 \times 10^{-4}$~M$_\odot$~yr$^{-1}$ 
and $\sim 3.7 \times 10^{-5}$~M$_\odot$~yr$^{-1}$ at $r \sim 2000$~AU 
for A and B, respectively. 
Even if a warmer initial gas temperature 
($\sim 20$~K) than in IRAM~04191 and some initial level of turbulence are
accounted for (see Di Francesco et al. 2001), these values of
$\dot{M}_\mathrm{inf}$ correspond to more than $\sim 15$ and $\sim 6$ times 
the canonical $a_\mathrm{eff}^3/G$ value, respectively (where 
$a_\mathrm{eff} \simlt 0.3\, \kms$ is the effective sound speed).  
These very high infall rates result both from very dense 
envelopes (see Motte \& Andr\'e 2001) and large, supersonic infall 
velocities ($\sim 0.68\, \kms $ and $\sim 0.47\, \kms $ for A and B, 
respectively, according to Di Francesco et al. 2001). 
Evidence of fast rotation, 
producing a velocity gradient as high as $\sim 40\, \kms \, $pc$^{-1}$,
was also found by Di Francesco et al. (2001) toward IRAS~4A.

\section{Conclusions: Comparison with collapse models}

In the case of low-mass, isolated dense cores, 
the SIS model of Shu~(1977) describes global features of the collapse 
(e.g. the mass infall rate) reasonably well and thus remains a useful,
approximate guide. 
In detail, however, the extended infall velocity profiles observed in 
prestellar cores (e.g. Lee et al. 2001) and in the very young Class~0 object
IRAM~04191 (\S ~3.2 above) are inconsistent with the pure inside-out collapse
picture of Shu (1977). The {\it shape} of the density profiles observed in 
prestellar cores are well fitted by purely thermal Bonnor-Ebert sphere models, 
but the {\it absolute} values of the densities are suggestive of some 
additional magnetic support (\S ~2.2). The observed infall velocities are
also marginally consistent with isothermal collapse models starting from 
%marginally stable equilibrium
Bonnor-Ebert spheres (e.g. Foster \& Chevalier 1993, Hennebelle et al. 2002),
as such models tend to produce somewhat faster velocities.
This suggests that the collapse of `isolated' cores is essentially 
{\it spontaneous} and somehow moderated by magnetic effects in magnetized, 
probably not strictly isothermal Bonnor-Ebert cloudlets. 
Furthermore, the contrast seen in Fig.~\ref{iram04191_model} between a 
steeply declining rotation velocity profile and a flat infall velocity 
profile beyond $\sim 3500$~AU suggests that angular momentum is {\it not}
conserved in the outer envelope of IRAM~04191. Such a behavior is very 
difficult to explain in the context of non-magnetic collapse models. 
In the presence
of a relatively strong ($\sim 60\, \mu$G) magnetic field, on the other hand,
the outer envelope can be coupled to, and spun down by, the (large moment 
of inertia of the) ambient cloud (e.g. Basu \& Mouschovias 1994). 
Based on a qualitative comparison with the ambipolar diffusion 
models of Basu \& Mouschovias, Belloche et al. (2002) propose that 
the rapidly rotating inner envelope of IRAM~04191 corresponds to a 
magnetically supercritical core decoupling from an environment still 
supported by magnetic fields and strongly affected by magnetic braking.
(Quantitatively, however, the models published so far are 
rotating too slowly to fit the observations.)
In this view, the inner $\sim 3500$~AU radius envelope of IRAM~04191 
would correspond to 
the effective mass reservoir ($\sim 0.5\, M_\odot $) from which the central 
star is being built. Moreover, comparison of these results with the 
rotational characteristics of other 
%(prestellar and protostellar) 
objects in Taurus (Ohashi et al. 1997) suggests that IRAM~04191 behaves 
in a typical manner and is simply observed particularly soon after point mass 
formation (i.e., at $t \simgt 0$). 
If this is correct, {\it the masses of the stars forming in 
clouds such as Taurus may be largely determined by magnetic decoupling effects}.

\begin{figure}[ht]
%\resizebox{\hsize}{!}{\includegraphics{/home/storage/pandre/tmp3/waterloo/phen_%waterloo.ps}}
\resizebox{\hsize}{!}{\includegraphics{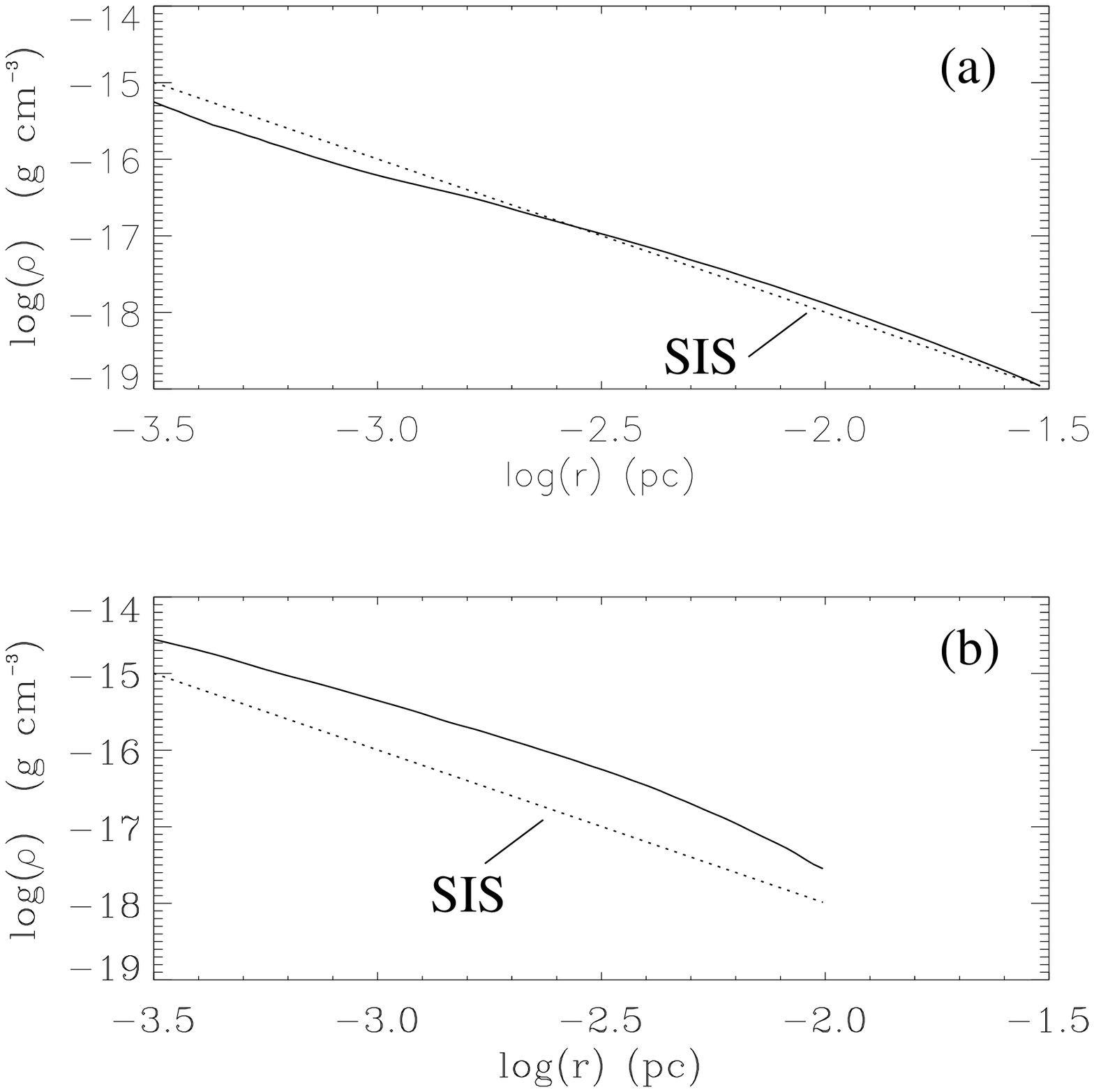}}
\caption{Density profiles (solid curves) obtained slightly after point 
mass formation ($t \simlt 10^4$~yr) in SPH numerical simulations of the 
collapse of a nearly critical [$(\rho_c/\rho_{out})_{init} \sim 12$]
%initially stable 
isothermal ($T= 10$~K) Bonnor-Ebert sphere induced by external 
compression (Hennebelle et al. 2002). For comparison, the dotted
line shows the $\rho \propto r^{-2}$ profile of a SIS at 10~K. 
In (a), the collapse was initiated quasi-statically (by very slow
compression with $P_{ext}/\dot{P}_{ext} = 20\ \times$ the initial
sound crossing time $R_{init}/\as $) 
and the density profile at $t \simgt 0$ is very similar to that of 
the SIS. By contrast, in (b), the collapse was induced by a very rapid increase
in external pressure (with $P_{ext}/\dot{P}_{ext} = 0.03 \times R_{init}/\as$), resulting in much larger densities around $t \sim 0$.
\label{hennebelle}}
\end{figure}

In protoclusters, by contrast, the large overdensity factors measured for
Class~0 envelopes compared to hydrostatic isothermal structures (cf. \S ~2.4),
as well as the fast supersonic infall velocities and very large infall
rates observed in some cases (e.g. \S ~3.2), are inconsistent with 
self-initiated forms of collapse and require a {\it strong external influence}.
This point is illustrated in Fig.~\ref{hennebelle} with the results of recent
SPH simulations by Hennebelle et al. (2002). These simulations 
follow the evolution of a Bonnor-Ebert sphere whose collapse
has been induced by an increase in external pressure $P_{ext}$. 
Large overdensity factors (compared to a SIS), together with supersonic 
infall velocities, and large infall rates ($\simgt 10\, \as^3/G$) 
are found near $t = 0$ when (and only when) the increase in $P_{ext}$ 
is strong and very rapid (e.g. Fig.~\ref{hennebelle}b), 
resulting in a violent compression wave. Such a violent collapse may be 
conducive to the formation of both massive stars (through higher accretion
rates) and multiple systems (when realistic, non-isotropic compressions are
considered). Future high-resolution studies with the next generation 
of (sub)millimeter instruments (e.g., ALMA) will greatly help test this
view and shed further light on the physics of collapse in cluster-forming regions.

\end{document}